**УДК 37.046.16+81’42**

**МОІСЕЄНКО Н. В.**,
кандидат фізико-математичних наук,
доцент кафедри інформатики та
прикладної математики
ДВНЗ »Криворізький національний
університет»

**МОІСЕЄНКО М. В.**,
асистент кафедри інформатики та
прикладної математики
ДВНЗ »Криворізький національний
університет»

**СЕМЕРІКОВ С. О.,**
доктор педагогічних наук, професор, в. о.
завідувача кафедри інженерної педагогіки
та мовної підготовки ДВНЗ «Криворізький
національний університет»

# МОБІЛЬНЕ ІНФОРМАЦІЙНО-ОСВІТНЄ СЕРЕДОВИЩЕ ВИЩОГО НАВЧАЛЬНОГО ЗАКЛАДУ

*У статті висвітлено визначення поняття мобільного інформаційно-освітнього середовища вищого навчального закладу, спрямованого на задоволення освітньо-наукових потреб всіх користувачів у забезпеченні необхідними електронними ресурсами у будь-який час та у будь-якому місці.*

**Постановка проблеми.** В сучасному світі в умовах інформатизації суспільства та високого рівня конкуренції на ринку праці постає проблема підготовки фахівців до використання сучасних інформаційних і комунікаційних технологій. Сучасний вищій навчальний заклад повинен стати осередком інноваційної освіти з випереджальною підготовкою фахівців нового покоління. Особливо це стосується підготовки професійно компетентних педагогів, здатних легко адаптуватися в сучасному освітньому середовищі, бути конкурентоспроможним в умовах сучасного ринку праці.

**Аналіз останніх досліджень і публікацій.** В. Ю. Биков, Ю. В. Горошко, М. І. Жалдак, Н. В. Морзе, Ю. В. Триус та інші вітчизняні фахівці один із ефективних шляхів вирішення проблеми підвищення якості освіти вбачають у впровадженні мобільних інформаційно-комунікаційних технологій (ІКТ), що на сучасному етапі розвитку ІКТ стають технологічною основою фундаменталізації навчання та створюють умови для реалізації мобільного навчання – сучасного напряму розвитку дистанційного навчання із застосуванням мобільних телефонів, смартфонів, КПК, електронних книжок та інших засобів. До головних переваг мобільного навчання відносяться: можливість навчання будь-де і будь-коли; особистісна зорієнтованість, портативність і мобільність засобів навчання; висока інтерактивність навчання; розвинені засоби спільної роботи; можливість безперервного доступу до навчальних матеріалів.

Впровадження в освіту ІКТ як сукупності «методів, засобів і прийомів, використовуваних для збирання, систематизації, зберігання, опрацювання, передавання, подання всеможливих повідомлень і даних» [1, с. 8], привело до виникнення терміну *інформаційно-освітнє середовище* (ІОС). Основними напрямками використання нових інформаційних технологій вважаються: 1) управління вищим навчальним закладом; 2) навчальний процес; 3) наукові дослідження.

ІОС вищого навчального закладу – системно організована сукупність засобів передавання інформації, принципів взаємодії учасників навчального процесу, дидактичного, організаційного та методичного забезпечення, яка орієнтована на задоволення потреб тих, хто навчається.

ІОС на базі інформаційно-комунікаційних технологій – системно організована сукупність засобів передавання даних, інформаційних ресурсів, протоколів взаємодії, апаратно-програмного, організаційного та методичного забезпечення, що орієнтована на задоволення потреб користувачів.

Основними функціями ІОС є:

− інформаційна (задоволення інформаційно-освітніх потреб учасників навчального процесу);

− освітня (своєчасне та якісне забезпечення навчальним матеріалом);

– контролююча (контролювання самостійної роботи слухачів);

− організаційна (взаємозв’язок з викладачами).

Розвиток ІОС охоплює велике поле застосування ІКТ в освіті. Перший напрямок, це забезпечення навчального процесу – лекційних, лабораторних та практичних занять, забезпечення розробки підручників і навчальних посібників, створення інформаційно-довідкової бази даних, підтримка дистанційного доступу до освітніх ресурсів. Застосування ІКТ дає можливість використовувати більш ефективні способи подання навчального матеріалу під час лекційних занять, підвищити якість засвоєння матеріалу шляхом використання різноманітних навчальних програм [2].

ІОС є сукупністю умов, що забезпечують діяльність та інформаційну взаємодію з розподіленими інформаційними ресурсами та користувачами, на основі сучасних інтерактивних засобів інформаційних і комунікаційних технологій, що орієнтовані на формування високоосвіченої й духовно розвиненої особистості, здатної до соціалізації в сучасних умовах.

ІОС середовище прямо чи опосередковано впливає на процеси становлення особистості в сучасних умовах. Отже, воно виконує інформативну та комунікативну функції; а також сприяє реалізації тих видів діяльності, що пов’язані з використанням комп’ютера та засобами ІКТ.

О. О. Андрєєв розглядає ІОС як педагогічну систему з підсистемами її забезпечення (фінансово-економічна, матеріально технічна, нормативно-правова, маркетингова та менеджменту) [3], тобто розглядає управлінську складову цього середовища.

Є. К. Марченко визначає ІОС як системно організовану сукупність освітніх закладів та органів управління, локальних та глобальних інформаційних мереж, книжкових фондів бібліотек, систему їх функціональної та територіальної адресації, нормативних документів, а також сукупність засобів передачі даних, інформаційних ресурсів, апаратно-програмного і організаційно-методичного забезпечення, які реалізують освітню діяльність [4], тобто розглядає технічну складову цього середовища.

О. О. Ільченко визначає ІОС як системно організовану сукупність інформаційного, технічного, навчально-методичного забезпечення, яка нерозривно пов'язана з людиною, як з суб'єктом освітнього процесу, тобто підкреслює зв'язок системи з розвитком особистості [5].

Всі три визначення мають спільну особливість: інформаційно-освітньому середовищу надаються системні властивості.

Втім, єдиного підходу до розуміння поняття ІОС стосовно до ВНЗ поки що не сформовано, авторами [6] пропонується використовувати таке означення: «інформаційно-освітнє середовище ВНЗ – педагогічна система, що об'єднує в собі інформаційні освітні ресурси, комп'ютерні засоби навчання, засоби управління освітнім процесом, педагогічні прийоми, методи та технології, спрямовані на формування інтелектуально розвиненої соціально-значущої творчої особистості, що володіє необхідним рівнем професійних знань та компетенцій». Л. Ф. Панченко, узагальнивши різні трактування ІОС, розглядає ІОС ВНЗ як відкриту багатовимірну педагогічну реальність, що включає психолого-педагогічні умови, сучасні інформаційно-комунікаційні технології і засоби навчання, і забезпечує взаємодію, співпрацю, розвиток особистості викладачів і студентів у процесі вирішення освітніх завдань [7, с. 78].

**Мета статті** – визначення поняття мобільного інформаційно-освітнього середовища вищого навчального закладу.

**Виклад основного матеріалу.** Закон про освіту Російської Федерації в редакції від 28.02.2012 включає в себе таке трактування (ст. 16, п. 3): електронне ІОС – це електронні інформаційні ресурси, електронні освітні ресурси, сукупність інформаційних технологій, телекомунікаційних технологій, відповідних технологічних засобів, що забезпечують засвоєння тими, хто навчається, освітніх програм в повному обсязі незалежно від місця знаходження.

Складові частини інформаційно-освітнього середовища можна умовно розділити на групи за типом користувачів, для яких вони призначені:

1. Загальна інформація: загальна інформація про навчальний заклад; інформація про освітні послуги, що надаються закладом, напрями навчання; інформація про студентське життя, новини тощо.

2. Навчально-методичні матеріали: освітні програми та навчальні плани спеціальностей; графік навчального процесу; робочі програми дисциплін; методичні матеріали (лекції, інструкції до виконання лабораторних робіт, плани семінарів тощо); інструктивно-методичні матеріали для виконання курсових та кваліфікаційних робіт; електронна бібліотека; навчальні та спеціалізовані програмні засоби (відповідно до спеціальності); система оцінки знань; системи спілкування між викладачами та студентами (форуми, електронна пошта); організаційні інформаційні повідомлення.

3. Адміністративна інформація: внутрішня документація університету, що не стосується навчального процесу; документація підрозділів (кафедри); документація деканатів (відомості про студентів, відвідування ними занять, електронні журнали успішності тощо).

На сучасному етапі розвитку ІКТ провідним способом доступу до складових ІОС університету є Інтернет-доступ через різні засоби мобільного навчання.

Автори [8] визначають мобільне навчання як етап еволюції електронного навчання. При цьому вони акцентують увагу на доступності інформації, що досягається за рахунок використання саме переносних пристроїв та адаптованих до них застосунків. Нажаль, при аналізі як електронного, так і мобільного навчання, дослідники зводять предмет дослідження до технічного забезпечення процесу навчання. Так, у [9] мобільне навчання описується як те, що відноситься «до застосування мобільних та портативних ІТ-пристроїв, таких, як кишенькові комп'ютери PDA (Personal Digital Assistants), мобільні телефони, ноутбуки та планшетні ПК у викладанні та навчанні».

Природно визначити мобільне навчання як навчання в мобільному ІОС, незалежне від місця знаходження того, хто навчається. При цьому мобільність ІОС не означає лише використання мобільних пристроїв, оскільки технічні пристрої не обумовлюють існування педагогічних систем, а впливають на їх розвиток. Технічною базою мобільного навчання можуть бути будь-які ІКТ, що дозволяють забезпечити мобільність робочого місця та його оточення, яке ми й називаємо мобільним ІОС.

За визначенням ЮНЕСКО, мобільний пристрій є цифровим, він легко переноситься, як правило, належить і контролюється індивідом, а не установою, може отримати доступ до Інтернет, має мультимедійні можливості, і може сприяти виконанню великої кількості завдань, зокрема, пов'язаних з комунікацією [10, с. 6]. Найчастіше мобільність Інтернет-пристрою пов'язується із здатністю людини до його переміщення у просторі без втрати доступу до послуг Інтернет. В. Ю. Биков наголошує, що сам мобільний пристрій, «як фізичний об'єкт неживої природи, звісно, не є і не може бути мобільним. Мобільним може бути лише Інтернет-користувач», оснащений ним [11, с. 23].

На думку В. Ю. Бикова, перспективним шляхом нівелювання різниці в оснащенні різних мобільних Інтернет-пристроїв є розвиток хмарної інфраструктури: «у найближчій перспективі вага і вартість МІП [мобільних Інтернет-пристроїв] мають бути суттєво знижені без втрати, навіть підвищення функціональності МІП щодо забезпечення ефективної ІК-діяльності користувачів» [11, с. 28]. Характеризуючи мобільні Інтернет-пристрої, В. Ю. Биков уводить поняття загального простору Інтернет-діяльності користувача, виділяючи у ньому підпростір Інтернет-доступності користувача (Інтернет-простір користувача), перебуваючи або переміщаючись з одного в інше місце в межах якого Інтернет-користувач може за певних умов (зокрема, використовуючи засіб Інтернет-доступу за наявності покриття простору Інтернет-сигналом) здійснювати інформаційно-комунікаційну діяльність [11, с. 14].

Простір Інтернет-доступності автор характеризує у термінах щільності (зокрема, оснащеності мобільними Інтернет-пристроями) та різноманітності (зокрема, за ступенем мобільності пристроїв) [11, с. 16-18]. Дослідник пропонує класифікувати засоби Інтернет-доступу за готовністю до Інтернет-застосування, за форм-фактором конструктивного виконання, за придатністю до переміщення тощо. Так, за ступенем придатності (пристосованості) засобів Інтернет-доступу до переміщення В. Ю. Биков поділяє їх на переносні (мобільні), пересувні та стаціонарні [11, с. 21-22]:

– переносний засіб Інтернет-доступу – пристрій індивідуального використання, форм-фактор якого (передусім, вимоги щодо масогабаритних та енергетичних параметрів пристрою) передбачає можливість для Інтернет-користувача переносити і використовувати такий пристрій в процесі здійснення власної ІК-діяльності;

– пересувний засіб Інтернет-доступу – пристрій як індивідуального, так і колективного використання, форм-фактор якого передбачає пристосованість такого пристрою до переміщення, в тому чисті у простір Інтернет-доступності. Такі засоби можуть потребувати для свого переміщення транспортних засобів (звичайних або спеціальних), а для придатності використання – також спеціальних засобів фіксації робочого положення. Це пристрої, що, окрім іншого, обов'язково має у своєму складі вхідні Інтернет-порти та інші комп'ютерні компоненти для опрацювання електронних даних;

– стаціонарний засіб Інтернет-доступу – пристрій як індивідуального, так і колективного використання, форм-фактор якого передбачає, що такий пристрій не змінює свого географічного розташування протягом тривалого часу і не пристосований до переміщення, в тому числі у простір Інтернет-доступності. Для забезпечення придатності використання такі засоби можуть потребувати спеціальних засобів фіксації робочого положення. Вони розміщуються як у різних за призначенням приміщеннях, так і поза ними (встановлюється за допомогою спеціальних постаментів на вулицях, площах, на зовнішніх і внутрішніх стінах будинків та ін.). Як і у випадку пересувних засобів Інтернет-доступу, стаціонарні обов'язково мають у своєму складі вхідні Інтернет-порти та інші комп'ютерні компоненти для опрацювання електронних даних.

При використанні мобільних Інтернет-пристроїв забезпечення необхідної насиченості простору Інтернет-доступності досягається лише за умови 100 % оснащеності користувачів ними – такий простір В. Ю. Биков називає мобільним. Мобільність простору одночасно визначається як рівнем Інтернет-доступності середовища (характеристиками техніко-технологічних умов забезпечення Інтернет-доступності простору – насиченості, складовими якої є щільність та різноманітність, і неперервності, складовими якої є територіальне і часове покриття Інтернет-сигналом), так і відповідними ІКТ-компетентностями Інтернет-користувача, що відображаються його характеристиками (властивостями). До останніх «варто, передусім, віднести таку особистісну характеристику Інтернет-користувача, як його навченість щодо ефективного використання ЗІД [засобів Інтернет-доступу] та Інтернет-технологій для здійснення ІК-діяльності в Інтернет-просторі (визначається сукупністю відповідних ІКТ-компетентностей користувача)» [11, с. 28] – мобільність Інтернет-користувача (мобільність користувача в просторі Інтернет-доступності).

Мобільний Інтернет-користувач на основі опанованих знань, умінь і навичок в ІКТ-сфері, сформованих відповідних ІКТ-компетентностей здійснює інформаційно-комунікаційну діяльність за допомогою засобів і технологій оточуючого його *мобільно орієнтованого середовища* – частини мобільного простору, комп'ютерно орієнтованого (комп'ютерно інтегрованого, персоніфікованого) відкритого середовища діяльності (освітньої, навчальної, управлінської та ін.) Інтернет-користувача, в якому створені необхідні і достатні умови для забезпечення його мобільності [11, с. 30]. Слід зазначити, що високий рівень мобільності Інтернет-користувача може бути досягнений і без використання мобільних Інтернет-пристроїв (за умови насичення простору пересувними та стаціонарними Інтернет-пристроями).

*Мобільно орієнтоване інформаційно-освітнє середовище вищого навчального закладу* визначимо як відкриту багатовимірну педагогічну систему, що включає психолого-педагогічні умови, мобільні інформаційно-комунікаційні технології і засоби навчання, наукових досліджень та управління освітою, і забезпечує взаємодію, співпрацю, розвиток особистості викладачів і студентів у процесі вирішення освітніх та наукових завдань у будь-який час та у будь-якому місці.

Дане трактування надає можливість узагальнити різні види мобільності (географічної, апаратної, навчальної, віртуальної [12], академічної [13] та ін. [14]) у межах одного середовища, що відображає одну із основ Болонського процесу та, як наголошується в комюніке «Простір європейської вищої освіти в новому десятиріччі», є одним з пріоритетів сучасної вищої освіти, разом з іншими перевагами:

– соціальний аспект: рівноправний доступ тих, хто навчається, до системи вищої освіти;

– безперервна освіта, що ґрунтується на принципі суспільної відповідальності;

– особистістно-орієнтоване навчання;

– взаємозв'язок освіти, досліджень та інновацій;

– академічна мобільність як фактор підвищення якості навчальних програм та досягнень в галузі наукових досліджень [15].

**Висновки.** Мобільне інформаційно-освітнє середовище ВНЗ забезпечує реалізацію ряду переваг: ефективного використання сучасних технічних засобів навчання; залучення кращих науково-педагогічних працівників; впровадження та підтримки авторських програм; забезпечення цілеспрямованого розвитку студентів. В умовах мобільного ІОС кожен студент має вільний доступ, незалежний від часу і місця, до будь-яких матеріалів з навчальних дисциплін, набуваючи при цьому необхідні для них практичні навички, реалізує корисну взаємодію, обмін знань, організовує безперервний процес навчання. Створення та підтримка мобільного інформаційно-освітнього середовища університету дозволить вивести діяльність вищого навчального закладу на якісно новий рівень та підвищити його конкурентоспроможність в сучасних умовах.

**Список використаної літератури**

**MOISEIENKO N.,**
Doctor of Philosophy (Physics), Associate Professor of Informatics and Applied Mathematics Department, SIHE «Kryvyi Rih National University»
**MOISEIENKO M.,**
Lecturer of Informatics and Applied Mathematics Department, SIHE «Kryvyi Rih National University»
**SEMERIKOV S.,**
Doctor of Science (Pedagogical Sciences), Professor, Acting Chair of Engineering Pedagogy and Language Training Department, SIHE «Kryvyi Rih National University»

## THE MOBILE INFORMATION AND EDUCATIONAL ENVIRONMENT OF HIGHER EDUCATIONAL INSTITUTION.

***Abstract. Introduction.*** *In the modern world in the conditions of informatization of society and high level of competition at the labor-market the problem of preparation of specialists appears to the* *use of modern information and of communication technologies. Modern higher educational establishment must become the core of innovative education with problem preparation of specialists of new generation. Especially it touches preparation professionally of competent teachers, capable easily to adapt oneself in a modern educational environment, be competitive in the conditions of modern labor-market.*

***Purpose.*** *To highlight the definition of mobile information and educational environment of higher educational institution.*

***Methods.*** *Theoretical analysis of existing points of view on the definition of similar environments.*

***Results.*** *Define the concept of mobile information and educational environment of a higher educational institution aiming to meet the educational and research needs of all users in providing the necessary e-resources anytime and anywhere.*

***Originality.*** *Theoretically grounded the concept of mobile information and educational environment of higher educational institution and its structure.*

***Conclusion.*** *Mobile information and educational environment of a higher educational institution ensures the realization of a few preferences: effective using modern technical learning tools; attracting the best educators; implementation and supporting author's courses; ensuring purposeful development of students. In such environment every student have free access (independent from time and place) to any materials from the academic disciplines, while gaining for them the necessary practical skills, useful implements interaction, knowledge sharing, organizes continuous learning process. The creation and support of mobile information and educational environment of higher educational institution will bring the University activities to a qualitatively new level and enhance its competitiveness in modern conditions.*